\documentclass[preprint,12pt]{elsarticle}

\usepackage{graphicx}

\usepackage{amssymb,amsmath}

\usepackage[nodots]{numcompress}

\begin{document}
\begin{frontmatter}


\title{An accordion lattice based on the Talbot effect}

\author{S. Deachapunya\corref{cor}}
\ead{sarayut@buu.ac.th}

\author{S. Srisuphaphon}
\cortext[cor]{Corresponding author}
\address{Department of Physics, Faculty of Science, Burapha university, Chonburi, 20131, Thailand}

\begin{abstract}
We introduce an idea of producing an optical lattice relied on the Talbot effect. Our alternative scheme is based on the interference of light behind a diffraction grating in the near-field regime. We demonstrate 1-D and 2-D optical lattices with the simulations and experiments. This Talbot optical lattice can be broadly used from quantum simulations to quantum information. The Talbot effect is usually used in lensless optical systems, therefore it provides small aberrations.
\end{abstract}

\begin{keyword}
Talbot effect; Optical diffraction; Diffraction gratings
\end{keyword}
\end{frontmatter}

\section{Introduction}

Optical lattices are of great importance in fields of quantum many body systems and even in quantum applications such as quantum information technology \cite{Bloch2005,Jaksch2004,Monroe2002}. An experiment of ultracold quantum gases in optical lattices is attractive for quantum investigations in fundamental physics since it provides a light-matter interaction in optical dipole forces \cite{Bernet2000} which can be both of low-field and high-field seeking states \cite{Bloch2004}. The interaction of bosonic atoms on the lattices, Bose-Hubbard model, can be realized in optical lattices \cite{Lewenstein2007}. Storing atoms in optical lattices can be used as a quantum register in a Mott insulator phase \cite{Greiner2002} which was suggested as a quantum computer \cite{Feynman1986,Porto2003,Jaksch2000,Prevedel2007}. Loading ultracold atoms of Bose-Einstein condensate (BEC) \cite{Davis1995,Anderson1995} in an optical lattice was studied \cite{Denschlag2002}. A study of ultracold atoms in an optical lattice with dynamically variable periodicity allows us to perform versatile experiments from quantum simulations with a small lattice spacing to quantum information with a large lattice spacing \cite{Al-Assam2010,Fallani2005}. Therefore, an optical lattice with real-time variable lattice spacings (accordion lattice) is worth for studying. Among various schemes to produce accordion lattices, the use of two parallel beams encountered each other at the focal plane of a lens is an excellent one \cite{Li2008,Williams2008}.

Here, we introduce an idea of producing an alternative optical lattice relied on the Talbot effect \cite{Talbot1836,Lohmann1971,Berry2001,Case2009}. There were some early proposed ideas and experiments of optical lattice produced by the Talbot effect but with a binary shape of the opening fraction (the ratio between the slit width and the grating period) $f=0.5$ \cite{Ovchinnikov2006,Sun2007a,Sun2007b}. We here study the Talbot optical lattice with various opening fractions and grating periods. We demonstrate our idea with the simulations and experiments. Talbot optical lattice is simple, and more stable. Also, the main setup is usually used in lensless optical systems, therefore it provides small aberrations.

\begin{figure*}
\centerline{\includegraphics[width=1.2\columnwidth]{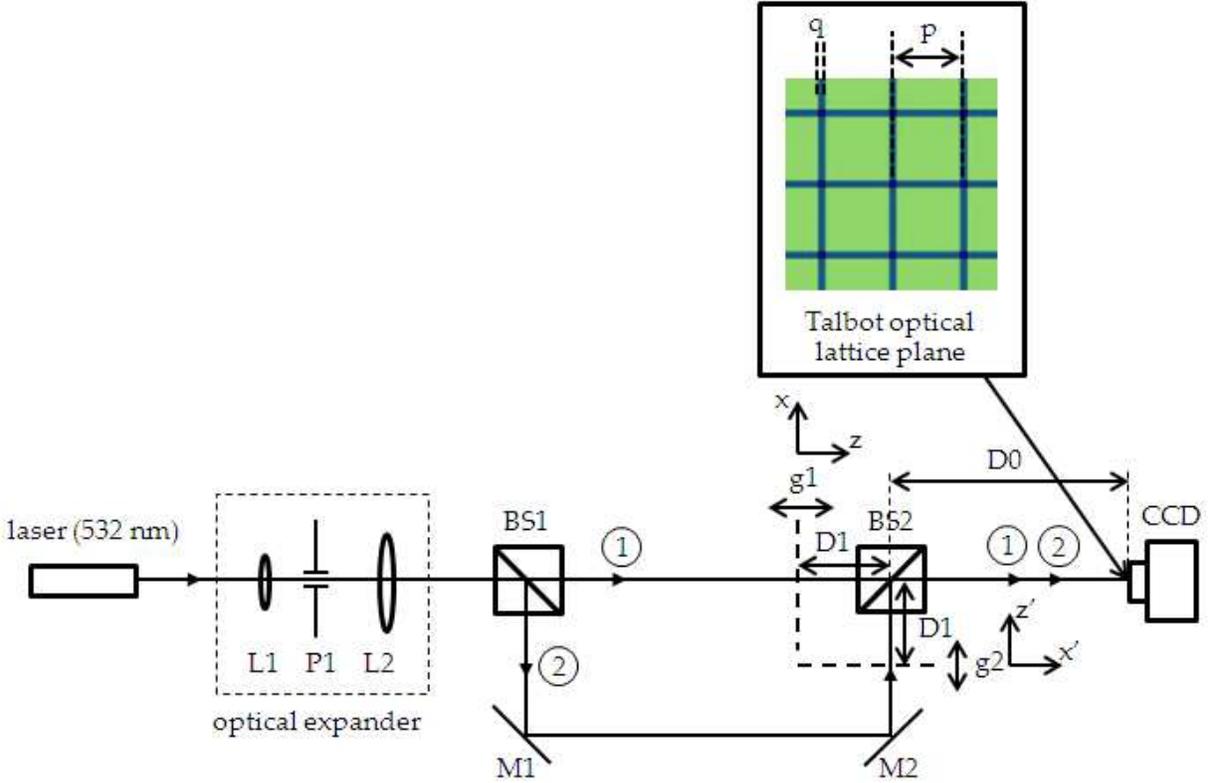}}
\caption{(Color online) Experimental setup of the real-time control optical lattice. L1, L2, P1 are convex lenses, and a pinhole as an optical expander; BS1 and BS2 are 50:50 beamsplitters; g1 and g2 are diffraction gratings; M1 and M2 are mirrors; and CCD for imaging the optical lattice. The inset shows a sketch of the 2-D optical lattice with g1 and g2 of $d=20\mu m,f=0.9$ and $z=z^{'}=L_{T}$ (p is the lattice spacing, and q is the width of the lattice). g1 and g2 are movable in order to change the length from each grating to the CCD.}
\label{Fig1}
\end{figure*}

\section{Theory and method}

\begin{figure}
\centerline{\includegraphics[width=0.6\columnwidth]{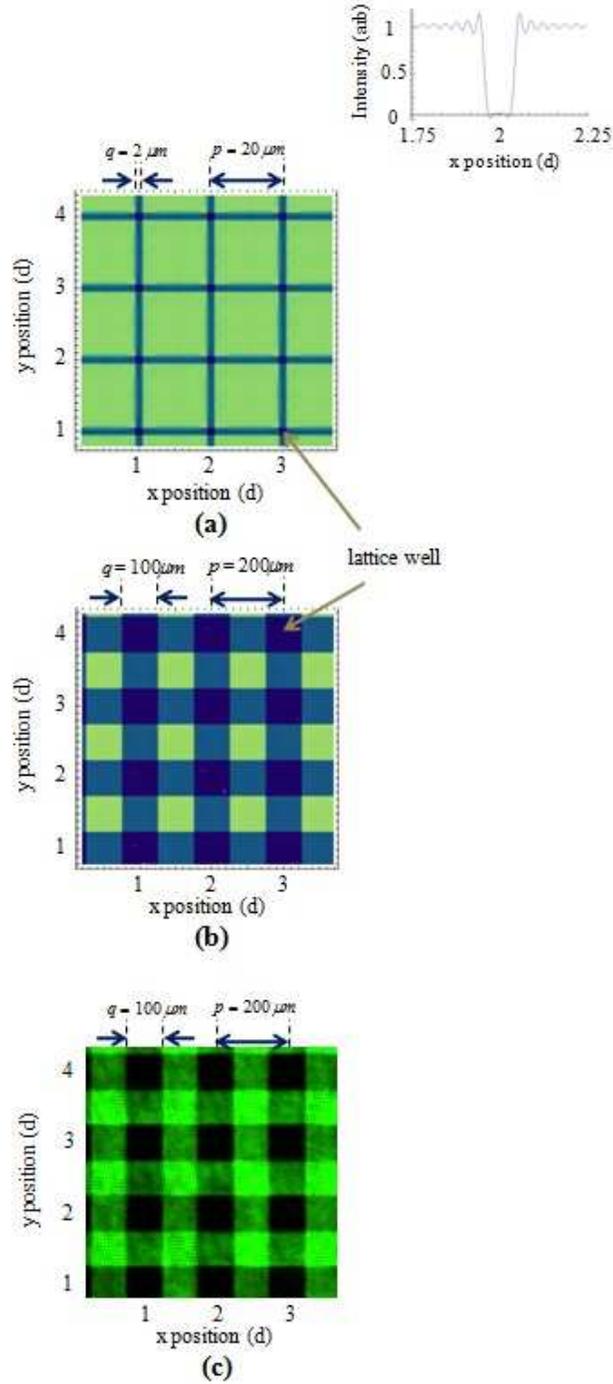}}
\caption{(Color online) The Talbot optical lattices: (a) Theoretical simulation according to Eq.~(\ref{Eq2}) with both gratings (g1 and g2) have the same period ($d=20\mu m,f=0.9$) and with $z=z^{'}=L_{T}$ (p=$d$). The inset shows a 1-D optical lattice with the same conditions but using only the first grating g1; (b) Theoretical simulation according to Eq.~(\ref{Eq2}) with both gratings have the same period ($d=200\mu m,f=0.5$) and with $z=z^{'}=L_{T}$; (c) Experimental lattice with the same conditions with (b). The lattice spacing (p) and width (q) represent in the unit of grating period ($d$).}
\label{Fig2}
\end{figure}

We show briefly the construction of optical lattice by using near-field diffraction pattern in the Talbot effect. Assuming a plane wave with the wave number $k$ propagating along $z$-axis falls onto a diffraction grating at $z=0$ (Fig.~\ref{Fig1}).  If the grating has the periodic modulation in the $x$ direction, the wave function propagates for a distance $z$ behind the grating can be expressed as \cite{Case2009}
\begin{equation} \label{Eq1}
 \psi(x,z)=\sum_{n}^{} A_{n}\exp\{ink_{d}x-\frac{in^2\pi z}{L_{T}}\},
\end{equation}
where $n=0,\pm 1,\pm 2,\dots$, $k_{d}=2\pi /d$, and $L_{T}=d^2/\lambda$ is the Talbot length.  Here, $d$ is the grating period, the coefficient $A_n= \sin(n \pi f)/(n \pi)$ represents the Fourier decomposition component of the periodic for the grating with an opening fraction $f$. The corresponding intensity $\psi^*(x,z) \psi(x,z)$  gives the near field interference pattern, so-called optical carpet \cite{Case2009}. This carpet can be considered as a 1-D optical lattice (Fig.~\ref{Fig2}(a) inset). Similarly, one can obtain the wave function also propagates along $z^{'}$-axis through the second grating (g2) but has diffraction pattern along $y^{'}$-axis by replacing $x$ in Eq.~(\ref{Eq1}) by $y^{'}$, i.e. $\psi(y^{'},z^{'})$ (Fig.~\ref{Fig1}). Therefore, we can produce a 2-D optical lattice that can be formed by the superposition of these two mentioned wave functions (Fig.~\ref{Fig1}). The intensity pattern produced by these combined waves can be obtained,
\begin{equation} \label{Eq2}
I= |\psi(x,z)+ \psi(y^{'},z^{'}) |^2.
\end{equation}

\section{Experiment}

Our 1-D Talbot optical lattice is produced with the 5 mW green diode laser ($\lambda=532nm$) as a coherent source which is expanded by an optical telescope to a diameter of 20 mm, and then illuminated a diffraction grating and downstream monitored by the imaging detection, i.e. CCD camera (DCU223C, Thorlabs). The telescope is used to expand the laser beam in order to make the homogeneous light amplitude distribution for the whole grating ~\cite{Salama1999} which makes the highest contrast of the Talbot image. The diffraction grating (chromium on glass, Edmund Optics inc., $d=200\mu m,f=0.5$) can be rotated with the rotation grating holder (LCRM2/M, Thorlabs) in order to align it to the camera axis. The distance from the first grating (g1) to the CCD (D0+D1) is set to be one Talbot length ($z=L_{T}=$75.2mm). Therefore, the lattice spacing (p) is equal to the grating period and the width (q) is half of the grating period for our demonstrated setup. The lattice depth can be estimated by measuring the intensity behind the grating with the photodiode (PM120D, Thorlabs).

The setup can be simply expanded to 2-D lattice by adding an additional grating (g2) as shown in Fig.~\ref{Fig1}. This second grating, which is similar to the first one, is placed and aligned perpendicular to the first grating (g1) where the grating lines of the first grating and second grating are perpendicular to each other or parallel to the $y$-axis and $x^{'}$-axis, respectively (Fig.~\ref{Fig1}). Then, the Talbot patterns are combined via the beamsplitter (BS2) and formed the 2-D optical lattice on the CCD plane (shown in the inset of Fig.~\ref{Fig1}). Each grating is placed above the translation stage (MTS50/M-Z8E, resolution 1.6 $\mu$m, Thorlabs) in order to adjust the length between the grating and the CCD (D0+D1). For the 2-D optical lattice, the setup can possibly be set with a 2-D cross grating but in this work we want to be able to demonstrate for both of 1-D and 2-D lattices.

\section{Results and discussion}

\begin{figure}
\centerline{\includegraphics[width=0.7\columnwidth]{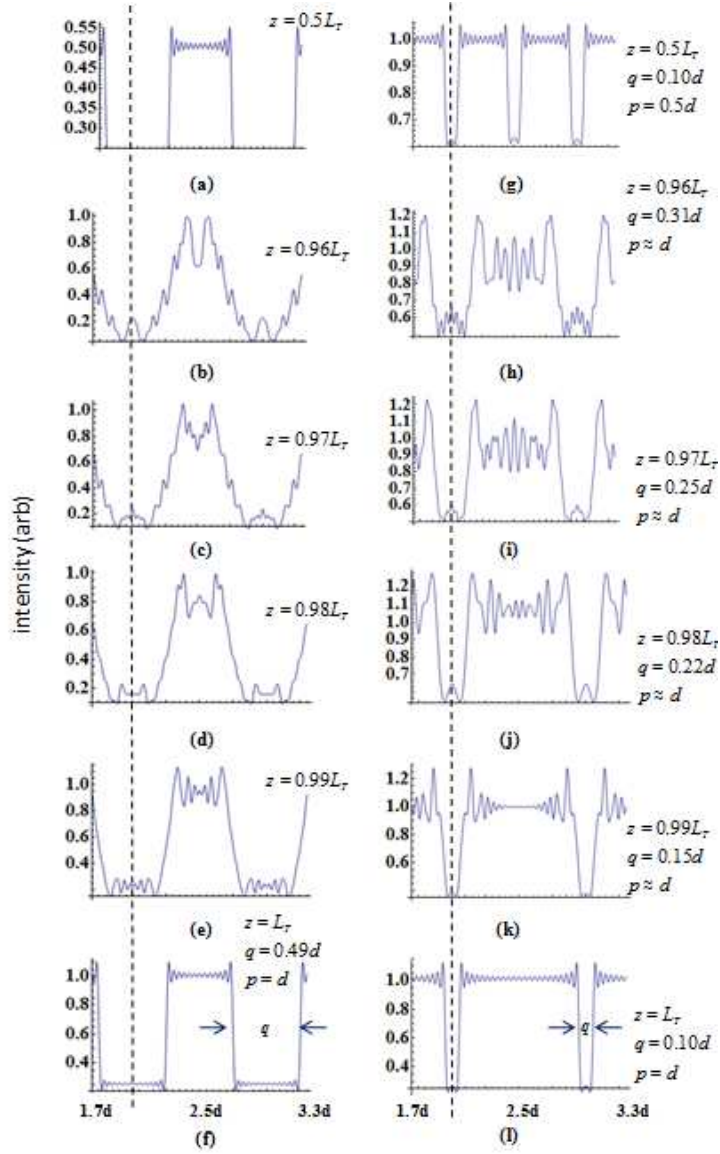}}
\caption{The cross section of theoretical 2-D optical lattices with both gratings have the same conditions of $d=20\mu m,f=0.5$ and $z$, $z^{'}$ varied symmetrically from $0.5L_{T}-1L_{T}$ (a)-(f); with $d=20\mu m,f=0.9$ and $z$, $z^{'}$ varied from $0.5L_{T}-1L_{T}$ (g)-(l). The horizontal axis represents the lattice spacing from 1.7$d$ to 3.3$d$. The vertical dash lines indicate the center of the lattices for each lattice site.}
\label{Fig3}
\end{figure}

We started with the simulations of the Talbot optical lattice. The 2-D optical lattices, simulated using Eq.~(\ref{Eq2}) are shown in Fig.~\ref{Fig2} (a) and (b). They were calculated with the truncated sum at $n=\pm 25$. The grating periods and opening fractions of $d=20\mu m,f=0.9$ and $d=200\mu m,f=0.5$, the laser wavelength of 532 nm, and the longitudinal distance $z=z^{'}=L_{T}$ were used in the calculations. The optical lattice shown in Fig.~\ref{Fig2} (a) has the square shape of 2$\mu$m width (q) with the lattice spacing (p) of 20$\mu$m which are corresponded to the grating periods $d=20\mu$m and opening fractions $f=0.9$ of the grating itself. The inset of Fig.~\ref{Fig2} (a) represents a 1-D optical lattice with the same conditions but using only the first grating g1. The lattice depth can be varied with the laser intensity and subsequently calculated to the potential depth. Also, Fig.~\ref{Fig2} (b) is the square optical lattice with the lattice spacing of 200$\mu$m and width of about 100$\mu$m since the gratings have $f=0.5$. We proved our idea by the experiment as shown in Fig.~\ref{Fig1} and the result shown in Fig.~\ref{Fig2} (c). Fig.~\ref{Fig2} (b) and (c) are the calculation and experiment with the same conditions. They show the similar shapes of the optical lattice. Fig.~\ref{Fig3} shows the cross section of 2-D optical lattices. The continuous change of the lattice spacings and lattice widths can be done by changing the longitudinal distances $z$ and $z^{'}$ for both gratings according to the fractional Talbot effect \cite{Case2009}. The $z$ and $z^{'}$ distances were varied equally from $0.5L_{T}$ to $1L_{T}$ ($d=20\mu m,f=0.9$) (Fig.~\ref{Fig3} (g)-(l)). The changes of $z$ and $z^{'}$ from $0.96L_{T}$ to $1L_{T}$ are corresponded to the lattice width of $q=0.31d=6.2\mu$m to $q=0.10d=2.0\mu$m (Fig.~\ref{Fig3} (h)-(l)). The width can also be $q=0.49d=9.8\mu$m at $z=z^{'}=0.9L_{T}$ (not shown in the figure). Therefore, the width can be varied from $2.0\mu$m to $9.8\mu$m with the small distance of about 75$\mu$m ($1L_{T}$ to $0.9L_{T}$). The lattice spacing is also changed to half of the grating period at $z=z^{'}=0.5L_{T}$. In contrast to this result, the condition of $d=20\mu m,f=0.5$ gives almost unchanged lattice width and spacing when $z$ and $z^{'}$ varied equally from $0.5L_{T}$ to $1L_{T}$ (Fig.~\ref{Fig3} (a)-(f)). These results show that the real-time control of the lattice width and spacing can be done with an asymmetrical grating, that is to say, $f=0.9$. In addition, the center of the lattice is stable and has no significant translation from the center within this small distance.

\section{Conclusions}

In conclusions, a 1-D and 2-D optical lattices with real-time control of the lattice spacing and width have been studied in our simulations and experiments. The optical lattice is produced by the Talbot effect. The lattice spacing and width can be changed with the longitudinal distances ($z$ and $z^{'}$) due to the Talbot and fractional Talbot effect. The results show that a symmetrical grating such as the opening fraction of 0.5 is not possible for changing the lattice spacing and width within a short distance, whereas an asymmetrical grating ($f=0.9$) provides the control change of $2.0\mu$m to $9.8\mu$m with the small longitudinal distance of about 75$\mu$m. An optical lattice produced by the Talbot effect overcomes the two counter-propagating laser beams, i.e. the standing wave of light because the variation of the lattice spacing and width can be more flexible than $\lambda/2$. Since the lattice spacing and width of the Talbot optical lattice can be modified with the longitudinal distance which also respected to the grating period, one can obtain almost arbitrary lattice spacings and widths with this method. The system can easily be extended to 3-D lattice by addressing atoms in a different layer of various longitudinal distances inside the Talbot optical lattice. Atoms and molecules in periodic potentials are an excellent tool for studying quantum mechanics. Even hot molecules can be used in an experiment of light-matter interaction with periodic potential \cite{Gerlich2007}. The Talbot optical lattice can be ultimately employed to study quantum tunneling when the spacing is small and possible quantum register for the large spacing.

S.D. acknowledges the support grant from the office of the higher education commission, the Thailand research fund (TRF), and Faculty of Science, Burapha university under contract number MRG5380264 and the National Electronics and Computer Technology Center, National Science and Technology Development Agency and Industry/University Cooperative Research Center (I/UCRC) in HDD Component, the Faculty of Engineering, Khon Kaen University. We thank Nupan Kheaomaingam for useful discussions.







\end{document}